# Chapter 6 in "Chondrules: Records of Protoplanetary Disk Processes"
# Vapor-Melt Exchange – Constraints on Chondrite Formation Conditions and Processes

Denton S. Ebel[1], Conel M. O'D. Alexander[2], and Guy Libourel[3]

[1]Department of Earth and Planetary Sciences, American Museum of Natural History, New York
Department of Earth and Environmental Sciences, Columbia University, New York, NY
Graduate School and Graduate Center of City University of New York (debel@amnh.org)

[2]DTM, Carnegie Institution of Washington, 5241 Broad Branch Road, Washington DC, 20015, USA. (calexander@carnegiescience.edu)

[3]Université Côte d'Azur, Observatoire de la Côte d'Azur, CNRS, UMR 7293 Lagrange, Boulevard de l'Observatoire, CS34229,06304 Nice Cedex 4, France. (guylibourel@gmail.com)

**Abstract:** The bulk volatile contents of chondritic meteorites provide clues to their origins. Matrix and chondrules carry differing abundances of moderately volatile elements, with chondrules carrying a refractory signature. At the high temperatures of chondrule formation and the low pressures of the solar nebula, many elements, including Na and Fe, should have been volatile. Yet the evidence is that even at peak temperatures, at or near the liquidus, Na and Fe (as FeO and Fe-metal) were present in about their current abundances in molten chondrules. This seems to require very high solid densities during chondrule formation to prevent significant evaporation. Evaporation should also be accompanied by isotopic mass fractionation. Evidence from a wide range of isotopic systems indicates only slight isotopic mass fractionations of moderately volatile elements, further supporting high solid densities. However, olivine-rich, FeO-poor chondrules commonly have pyroxene-dominated outer zones that have been interpreted as the products of late condensation of $SiO_2$ into chondrule melts. Late condensation of more refractory $SiO_2$ is inconsistent with the apparent abundances of more volatile Na, FeO and Fe-metal in many chondrules. Despite significant recent experimental work bearing on this problem, the conditions under which chondrules behaved as open systems remain enigmatic.

## 6.1  Introduction

There has been a longstanding debate about whether chondrules behaved as open chemical systems, that is, gaining or losing material by exchange with surrounding $H_2$-rich vapor during their formation (Wood, 1996; Hewins and Zanda, 2012; Connolly and Jones, 2016). Alterna-tively, chondrule chemical and isotopic compositions may primarily record the compositions of their precursors (Chapter 2), and they remained essentially closed systems upon melting and solidification.

Experimental simulations of chondrule textures suggest that they were rapidly heated to near liquidus temperatures of 1500-1700°C for relatively brief periods and then cooled to solidus temperatures (1000-1200°C) at 10-1000°C/hr (Hewins et al., 2005; Chapter 3). This implies formation timescales of hours to days. On these timescales, interactions between chondrules and gas would have been inevitable. Here we review the evidence for such interactions, note the constraints that they place on formation conditions, and highlight some unanswered questions.

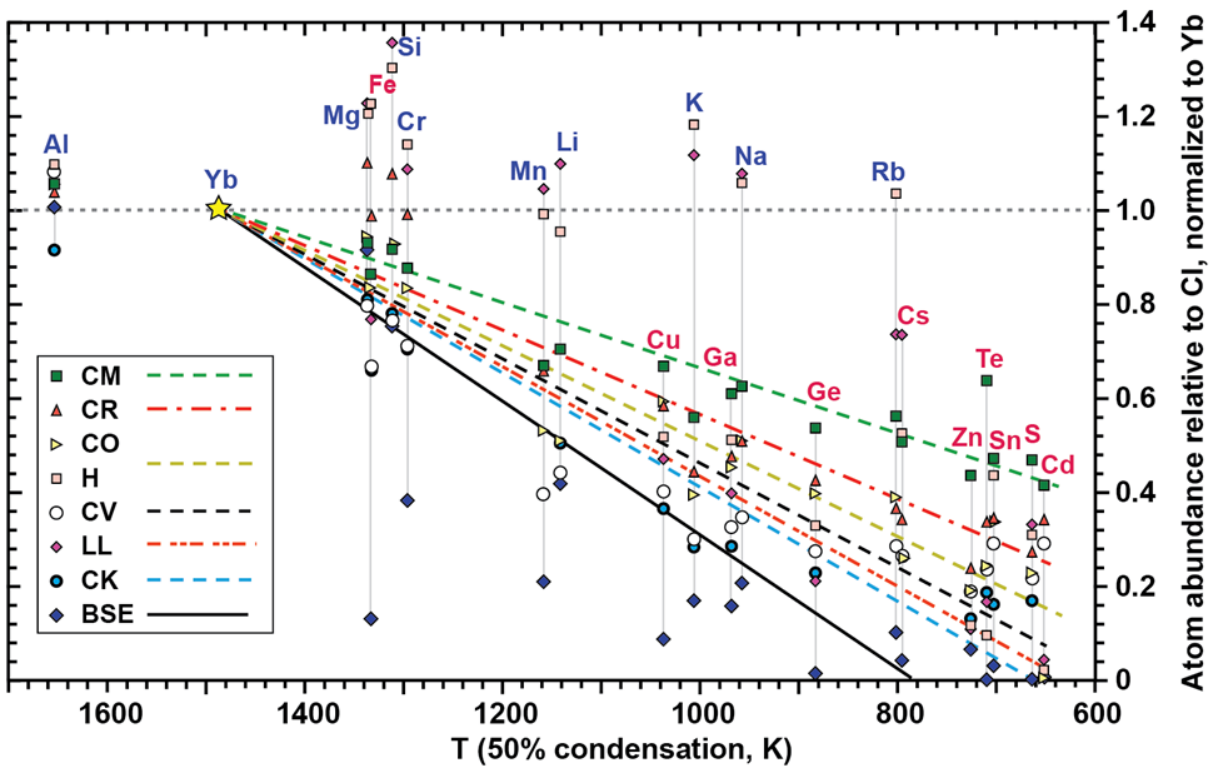

Figure 6.1. Volatile trends in chondrites and silicate Earth. Bulk silicate Earth (BSE, filled diamonds), and CK (filled circles), LL (open diamonds), CV (open circles), H (open squares), CO (filled triangles), CR (open triangles), CM (filled squares) chondrite atomic abundances of selected elements, normalized to Yb and CI chondrite (dotted line at '1') are plotted against their estimated 50% condensation temperatures (Lodders 2003). Earth data are from McDonough (2014); CI chondrite from Lodders et al. (2009); CK, LL, CV, H, CO, and CM from Wasson and Kallemeyn (1988); CR from Lodders and Fegley (1998). Trend lines for BSE and chondrites are calculated as described in text. Trends for CO and H are coincident. T(50%) for Rb and Cs are offset 20 K for clarity in this diagram.

## 6.2 Clues from the Bulk Volatile Contents of Chondrites

A starting point in the discussion of volatile elements is the relationship between chondrules and the bulk volatile content of their host meteorites, and by extension their parent planetesimals. It has been concluded (Lodders et al., 2009) that deviations of the chondrite groups from CI composition can be understood, "at least in principle, by gas-solid fractionation processes". Figure 6.1 illustrates the volatile element depletion of the bulk silicate Earth, known from the lithophile element abundances in mantle-derived rocks, compared to the volatile element depletions in LL and H ordinary chondrites (OC), CM, CR, CO, CV, CK, CH carbonaceous chondrites (CC), and EH enstatite chondrites. Elements are ordered according to their estimated 50% condensation temperatures (Lodders 2003; Table 6.1). Abundances are normalized to the CI carbonaceous chondrite standard (Lodders et al., 2009) and to the refractory lithophile element Yb, as this is well measured in CI chondrites (Lodders et al., 2009), and, in contrast to normalization to Mg, reveals depletions in Fe, Mg and Si. Furthermore, Yb is highly correlated with Al, often used as a proxy for refractory elements in general (Figure 6.1). Data for Mg, Fe, Si, Cr, Mn, Li, Cu, K, Ga, Na, Ge, Rb, Cs, Zn, Te, Sn S, and Cd are plotted. Recent analytical results are consistent with the older data used in Figure 6.1. For example, most analyses of Allende (CV3) by Makashima and Nakamura (2006) are within the standard deviation of 37 analyses by Stracke et al. (2012) for a majority of elements, and close to those of Jarosewich et al. (1987). The volatile trend lines were regressed using all these elements, except: Li in CR; Rb, Cs and Cd in CK; and lithophiles Mg, Si, Cr, Mn, Li, K, Na, Rb, and Cs in H and LL. The regressions (Table 6.2, Figure 6.1) were constrained to pass through Yb. In the LL and H chondrites, the lithophile elements are enriched, but only weakly collinear. The EH and EL enstatite chondrites are omitted because their highly reduced nature indicates formation of many of their components in reducing conditions, which would strongly affect the condensation temperatures of moderately volatile elements (Lodders, 2003; Ebel and Alexander, 2011; Ebel and Sack, 2013).

The depletion patterns in Figure 6.1 have been and continue to be the basis for extensive arguments about chondrule formation, summarized by Wood (1996). A two-component model mixing devolatilized chondrules and metal with CI-like matrix (e.g., Anders, 1964, 1977) would produce a step function, not continuous trends. Wood (1996) noted that a fractional condensation model (e.g., Wai and Wasson, 1977; Wasson, 1977) involving "systematic withdrawal of gas containing uncondensed volatile elements as the protochondritic system cooled" could produce these trends, and that it is "a wonder" that such good correlations can be produced. However, Wood also noted that for the elements with T(50% condensation) below about 700 K, the trends flatten. Indeed, these two models, and the fate of the missing volatiles, comprise 13-15 of the 15 "unresolved issues" in chondrule formation identified by Wood (1996). While many solutions have been proposed, their resolution based on meteoritic evidence remains elusive.

| symbol | host | T50%K |
|--------|------|-------|
| Al | hibonite | 1653 |
| Mg | forsterite | 1336 |
| Fe | Fe alloy | 1334 |
| Si | fo + en | 1310 |
| Cr | Fe alloy | 1296 |
| Mn | fo + en | 1158 |
| Li | fo + en | 1142 |
| Cu | Fe alloy | 1037 |
| K | feldspar | 1006 |
| Ga | Fe alloy+fsp | 968 |
| Na | feldspar | 958 |
| Ge | Fe alloy | 883 |
| Rb | feldspar | 800 |
| Cs | feldspar | 799 |
| Zn | fo + en | 726 |
| Te | Fe alloy | 709 |
| Sn | Fe alloy | 704 |
| S | troilite | 664 |
| Cd | en + troilite | 652 |

Table 6.1: Elements, primary host phases, and 50% condensation temperatures (K; Lodders, 2003) used to construct Figure 6.1. Abbreviations: fo=forsterite, en=enstatite, fsp=feldspar.

    Grossman (1996) attributed the fractionated bulk chondrite compositions in Figure 6.1 to losses or gains of at least six groups of elements. He concluded that these fractionations occurred "in chondrite-formation regions before chondrules" were formed. Similarly, Bland et al. (2005) found that although carbonaceous chondrites show monotonic volatile depletions in bulk, this pattern is not observed for matrix analyses. They attributed exchange of volatile elements between chondrules and matrix to thermal processing during subsequent chondrule formation, i.e., there is complementarity between matrix and chondrules (Bland et al., 2005; Palme et al., 2014; Ebel et al., 2016; Chapter 4). However, Zanda et al. (Chapter 5) contend that the matrices in the least altered carbonaceous chondrites are relatively unfractionated, and attribute the fractionations reported by Bland et al. (2005) to parent body processes. At present, the debate about whether chondrite matrices are dominated by a uniform, primitive CI-like material or material from the same reservoir from which chemically complementary chondrules formed remains unresolved, although isotopic evidence for the latter is accumulating (Chapter 10).

| type | slope | r squared | matrix |
|------|-------|-----------|--------|
| CI   | 0     |           | 0.998  |
| CM   | 0.00068 | 0.98    | 0.7    |
| CO   | 0.00100 | 0.98    | 0.419  |
| CV   | 0.00107 | 0.96    | 0.4502 |
| CR   | 0.00089 | 0.96    | 0.38   |
| LL*  | 0.00115 | 0.99    | 0.25   |
| H*   | 0.00102 | 0.94    | 0.23   |
| CK   | 0.00119 | 0.97    | 0.4    |
| CH   | 0.00109 | 0.77    | 0.05   |
| BSE  | 0.00139 | 0.33    |        |

Table 6.2: Results of the regressions computed for T(50%) vs. depletion, producing the lines shown in Figure 6.1, with calculated r-squared. Fraction of matrix used in Figure 6.2 is noted (sources in text). * see text for elements included in regressions for H and LL.

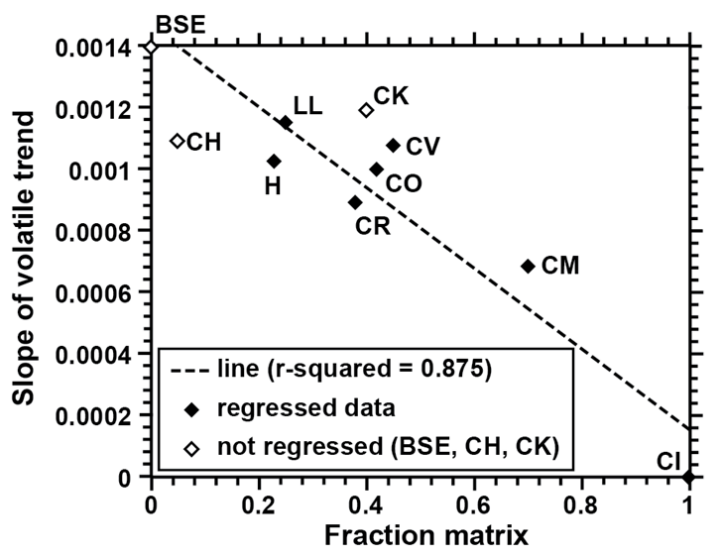

Figure 6.2. Relationship between slopes of volatile trends and matrix abundances (Table 6.2). Slopes (Table 6.2) are those of the lines in Figure 6.1. Sources of matrix data are described in the text.

There is a correlation between the volatile depletions of chondrites and their volume fractions of matrix. This correlation is discernable in the regressed slopes of the volatile trend lines in Figure 6.1, against the volume fraction of matrix (Figure 6.2; $R^2$=0.975). Matrix fractions (Table 6.2) are from Weisberg et al. (2006) and more recent measurements (Lobo et al., 2014, LL; Bayron et al., 2014, CR; and Ebel et al., 2016, CO and CV). The correlation is consistent with roughly CI abundances of the highly volatile elements, with T(50%) < ~750 K, in the matrix. In support of a very primitive, unheated component in matrix are the CI-like matrix-normalized abundances of presolar grains, volatile elements like Zn and organic C in the least metamorphosed chondrites (Huss 1990; Huss and Lewis, 1995; Alexander 2005; Davidson et al. 2014). Neither the presolar grains (or the noble gases they contain) nor the organic C would survive sig-nificant heating, let alone to temperatures required for chondrule formation (Hubbard and Ebel, 2015).

This observed correlation between chondrule abundance and volatile depletion strongly suggests that chondrules are the carriers of volatile depletion in meteorite parent bodies. Furthermore, volatile element systematics suggest that planet formation and chondrule formation are related processes. Although some objects in CB and/or CH chondrites may have impact origins, there are a great many arguments against such origins for the vast bulk of chondrules, including complementarity between matrix and chondrules (Bland et al., 2005; Palme et al., 2014; Chapter 4), among chondrules themselves (Ebel et al., 2008; Jones, 2012), and the ubiquity of chondrules among primitive materials (Taylor et al., 1983; Grossman, 1988; Wood, 1996). It may be significant that bulk silicate Earth (BSE) plots at approximately "zero matrix" in Figure 6.2.

### 6.3 Measurements of Chondrules

At low pressures and chondrule liquidus temperatures (1800-2100 K) many elements are predicted to be volatile based on thermochemical calculations, and this has been confirmed by numerous experiments (e.g., Hashimoto, 1983; Davis et al., 1990; Floss et al., 1996; Wang et al., 2001; Yu et al. 2003; Richter et al., 2011). The presence of $H_2$ enhances evaporation rates above pressures of $>10^{-6}$ bars (Nagahara and Ozawa, 1996; Kuroda and Hashimoto, 2002; Richter et al., 2002). For major elements, the order of decreasing volatility is S>alkalis>Fe>Si>Mg. Both theory and experiments indicate that the timescales for significant evaporation to occur at near peak chondrule formation temperatures would have been much shorter than the timescales for chondrule formation inferred from simulations of chondrule textures (Hewins et al., 2005). During free evaporation (i.e., no back reaction of a melt with evaporated material, and diffusion in the melt that is faster than evaporation), elemental fractionation is accompanied by isotopic fractionation with the melt becoming increasingly enriched in heavy isotopes relative to lighter ones as the concentration of the element decreases (reviewed by Davis et al., 2005 and Day and Moynier, 2014). The relationship between heavy isotope enrichment and element depletion follows the Rayleigh fractionation law, $R = R_0 f^{(\alpha-1)}$ in which the ratio of isotopes (e.g., $^{66}Zn/^{64}Zn$) remaining is, ideally, related the initial isotope ratio ($R_o$) and the fraction ($f$) of the element remaining (e.g., Zn) by the inverse square root of the mass ratios of the evaporating species (α, Davis and Richter, 2014).

Hence, if chondrules formed in the low pressure, low density environment typically invoked for protoplanetary disks, they would be expected to show the characteristic Rayleigh fractionation behavior of evaporation in many elements. As summarized below, there have been multiple searches for evidence for evaporation in chondrules (Table 6.3). However, at present there is no unequivocal evidence for large degrees of evaporation.

### 6.3.1 Sulfur

Despite its volatility, one study of S isotopes in Semarkona (LL3.00) chondrules (Tachibana and Huss, 2005) found no evidence for evaporation, which would mass-dependently fractionate $^{32}S$ from $^{34}S$, leaving heavy S in the melt. However, the petrologic evidence for S being present in chondrule melts during their formation is controversial. It is clear from experiments (Lauretta et al., 1997) that S will enter chondrules during parent body alteration/metamorphism, as observed in grades > 3.6 (Rubin et al., 1999), so careful study of unequilibrated chondrites is required. Hewins et al. (1997) argued that the most reduced (Type I) chondrules with coarsest phenocrysts lost more S than finer-grained, less thermally processed chondrules (cf., Zanda et al. 1994; Sears et al., 1996; Zanda, 2004). Rubin et al. (1999) found troilite (FeS) equally abundant in both highly reduced porphyritic and low-FeO chondrules in Semarkona (LL3.0), with no significant correlation between FeS content and grain size. The study of S is further hampered by the difficulty of accurately measuring the S content of Fe-Ni-S beads due to the heterogeneous exsolution of sulfides during crystallization and in the solid state. Despite this difficulty, Marrocchi and Libourel (2013) and Piani et al. (2016) have shown in CV and EH chondrites that sulfur concentrations and sulfide occurrence in chondrules obey high temperature sulfur solubility and saturation laws, and that gas-melt interactions with high partial vapor pressures of sulfur could explain the co-saturation of low-Ca pyroxene and troilite (cf., Lehner et al., 2013). Nevertheless, it remains important to resolve whether the S in chondrules is primary and, if so, whether it retains isotopic evidence for evaporation because S can potentially place the strictest constraints on chondrule formation of any major element.

### 6.3.2 Zinc, Cadmium and Copper

Zinc and Cd are even more volatile than S (Table 6.1). High-precision isotopic measurements of the five stable Zn isotopes in bulk carbonaceous and ordinary chondrites show that ratios such as $^{66}Zn/^{64}Zn$ are mass-dependently fractionated within chondritic materials (Luck et al., 2005; Moynier et al., 2009), but the Zn depletions in chondrites and their chondrules are not the result of evaporation (Figure 6.3; Moynier et al., 2017). Luck et al. (2005) measured whole rock $^{66}Zn/^{64}Zn$, $^{67}Zn/^{64}Zn$ and $^{68}Zn/^{64}Zn$ in a large suite of ordinary and carbonaceous chondrites. Recently, Pringle et al. (2017) confirmed and extended the previous measurements. Both found that bulk chondrites become isotopically *lighter* with decreasing Zn content (Figure 6.1). Were Zn depletion in chondrites due to evaporation, the Zn remaining in the solids would be expected to be enriched in heavier isotopes.

Luck et al. (2005) preformed sequential leaching of Krymka (LL3.1), and Pringle at al. (2017) measured Zn isotopes in magnetic, sulfide and silicate fractions separated from bulk chondrites Clovis (H3.6), GRA 95208 (H3.7) and ALH 90411 (L3.7). Sulfides were enriched in $\delta^{66}Zn$ by ~0.65 ‰ relative to silicates in Krymka, the same as the mean enrichment found by Pringle et al. (2017). Pringle et al. (2017) also measured $^{66}Zn/^{64}Zn$ and $^{68}Zn/^{64}Zn$ in Allende matrix and in chondrules separated from CV3 chondrites Allende and Mokoia, finding that chondrules are isotopically lighter than bulk Allende, and matrix is heavier, consistent with the results for Krymka.

Cadmium is a more volatile chalcophile metal than Zn (Figure 6.1; Wombacher et al., 2003, 2008). The $\delta^{114/110}Cd$ compositions of bulk Allende, Murchison (CM2) and Orgueil (CI) are identical to Earth's; however, ordinary and some enstatite chondrites show mass-dependent fractionations. Wombacher et al. (2008) found that Allende chondrules and Ca-, Al-rich inclusions (CAIs) were isotopically lighter in Cd than bulk, $\delta^{114/110}Cd$ = -0.1 ‰. They also found that an Allende sample, heated at 1100°C for 96 hours at the Ni-NiO oxygen buffer, became lighter, with $\delta^{114/110}Cd$ = -0.9 ‰, likely due to the survival of refractory host minerals (i.e., in chondrules and CAIs). Russell et al. (2003) measured $\delta^{65}Cu$ in NWA 801 (CR2) and reported that the chondrules are isotopically lighter than the bulk rock. Thus Cd and Cu isotopes appear to behave similarly to Zn isotopes, however it must be noted that these elements, and Zn, are highly mobile and may be isotopically fractionated during parent body aqueous alteration (Walsh and Lipschutz, 1982; Palme et al., 1988; Friedrich et al., 2003).

| element | T(50%) cond. | maximum fractionation | references |
|---------|-------------|----------------------|------------|
| Cd | 642 | chondrules are lighter than bulk | Wombacher et al., 2003, 2008 |
| S | 664 | < 1 ‰/amu | Tachibana and Huss, 2005 |
| Zn | 726 | chondrules are lighter than bulk | Luck et al., 2005; Pringle et al., 2017 |
| K | 1006 | < 1 ‰/amu | Alexander et al., 2000; Alexander and Grossman, 2005 |
| Fe | 1334 | < 1 ‰/amu | Alexander and Wang, 2001; Zhu et al. 2001; Kehm et al., 2003; Mullane et al., 2005; Poitrasson et al. 2005; Needham et al., 2009; Hezel et al., 2010; Wang, 2013 |
| Si | 1310 | ≤ 1.5 ‰/amu | Molini-Velsko et al., 1986; Clayton et al., 1991; Georg et al. 2007; Hezel et al., 2010; Armytage, 2011; Kühne et al., 2017 |
| Mg | 1336 | ~ 1 ‰/amu | Esat and Taylor, 1990; Galy et al., 2000; Young et al., 2002; Bouvier et al., 2013; Deng et al., 2017 |

Table 6.3: Isotopic fractionations observed in meteoritic chondrules (T(50%) condensation from Lodders, 2003).

Luck et al. (2005) interpreted their measurements as due to interaction of Zn-depleted refractory components with vapor enriched in isotopically light Zn. Pringle et al. (2017) went further, suggesting that evaporative loss from sulfide that was enriched in the heavy isotopes of Zn from chondrite-forming reservoirs caused both Zn elemental depletions and enrichment in light Zn isotopes. A similar explanation may apply to chalcophile Cu and Cd. It is of particular interest that the differing depletions of chalcophile and lithophile elements in ordinary chondrites (Figure 6.1) may be related to the isotopic fractionations of some of the chalcophile elements, as these authors suggest.

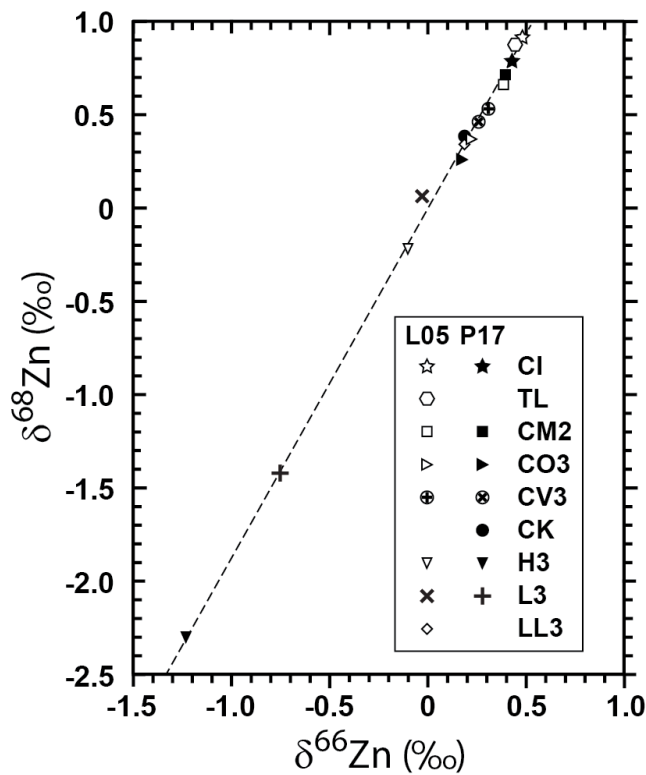

Figure 6.3. Three isotope plot of $\delta^{68}Zn$ versus $\delta^{66}Zn$ showing average values for bulk chondrites reported by Luck et al. (2005, "L05") and Pringle et al. (2017, "P17"). "TL" refers to the Tagish Lake meteorite. Dashed line (slope 1.86) is a regression of Pringle et al. (2017) average values with the mean CI value of Barrat et al. (2012, at 0.46, 0.88).

### 6.3.3 Alkali Metals: Potassium and Sodium

Sodium (Na) and potassium (K) are the next most volatile of the major/minor elements in chondrules, and their abundances in chondrule mesostases can vary by over an order of magnitude. Sodium is monoisotopic, but K has two abundant stable isotopes and is an obvious target for isotopic searches for evidence of evaporation (Humayun and Clayton, 1995). Richter et al. (2011) showed that K diffusion in chondrule liquids is sufficiently rapid that there would be no suppression of isotopic fractionation by diffusion-limited evaporation. Thus, if there was free evaporation of K, it should follow the expected Rayleigh fractionation law, as is observed in micrometeorites and cosmic spherules that were heated during atmospheric entry (Taylor et al. 2005). However, some caution must be taken when studying the alkalis because they can be very mobile in the glasses where they are concentrated, both during aqueous alteration and during thermal metamorphism. Thus, only chondrules from the most pristine chondrites should be studied, and even then considerable care is needed to only measure chondrules that do not appear to have exchanged with matrix (Huss et al., 2005).

Secondary ion mass spectrometric (SIMS) measurements of K isotopic compositions in Semarkona (LL3.0) chondrules revealed no systematic isotopic fractionations consistent with Rayleigh fractionation (Alexander and Grossman, 2005). There were some isotopic fractionations reported, but they were not systematic, and were attributed to instrumental artifacts associated with microcrystallites and fractures/holes in the mesostases that were measured. Alkali abundances are much lower in carbonaceous chondrite chondrules, suggesting that they could have large isotopic anomalies, but the low elemental abundances make the measurements very challenging. To date, no study of K isotopes in carbonaceous chondrite chondrules has been carried out.

One potential explanation for the lack of K isotopic fractionations in Semarkona chondrules is that the K condensed into the chondrules at relatively low temperatures shortly before or even after final solidification. Indeed, this is what thermochemical models predict should happen under some conditions (Ebel and Grossman, 2000). The measured distribution coefficients between glass and clinopyroxene crystallites in the mesostases of chondrules show that Na, at least, was present in the chondrules prior to final solidification of the glass (Jones, 1990, 1996; Libourel et al., 2003; Alexander and Grossman, 2005). However, it is still possible that the alkalis entered the chondrules just before growth of the clinopyroxene crystallites. One way to test this explanation is to look for alkali zoning in chondrule phenocrysts. The expectation would be that Na contents would be very low throughout most of the phenocrysts' growth, and rise very dramatically at their edges. Olivine is the most common phenocryst type, is normally the first phase to have begun crystallizing, and often exhibits igneous zoning in its major and minor elements. The olivine-melt distribution coefficient for Na is small (~0.003, Borisov et al., 2008). Nevertheless, it is high enough that Na concentrations in ordinary chondrite chondrule olivines can be measured by electron probe.

Chondrules with porphyritic textures are thought to have approached their liquidus temperatures without destruction of all crystal nuclei so that crystal growth would have begun soon after the onset of cooling (Hewins et al., 2005). Thus, their phenocrysts should preserve a record of conditions throughout a chondrule's cooling history. After carefully selecting only olivine-rich, porphyritic Semarkona chondrules with at least some interior regions of their mesostases that had not exchanged alkalis with the matrix, Alexander et al. (2008) measured the Na zoning profiles in their phenocrysts. Surprisingly, they found that contrary to expectations the phenocrysts had measurable levels of Na even in their cores and only modest increases toward their rims. Indeed, the zoning profiles were roughly consistent with essentially closed system crystallization. This was true for chondrules with wide ranges of Na and FeO contents, and liquidus temperatures. One obvious concern is that the zoning profiles are the result of later diffusion of Na into the olivine. However, Na diffuses relatively slowly in olivine. The remarkable conclusion from these observations is that Na was present in Semarkona chondrules at roughly the presently observed abundances throughout their crystallization. These measurements have been confirmed by several independent studies (Borisov et al., 2008; Hewins et al., 2012; Kropf and Pack, 2008; Kropf et al., 2009; Florentin et al., 2017), although there is some debate about exactly how much Na loss/gain (10-50 %) during chondrule formation and cooling is allowable.

Carbonaceous chondrite chondrules generally have much lower Na contents than ordinary chondrite chondrules, which is also reflected in the bulk chondrite compositions (Figure 6.1). Hence, the chondrules in carbonaceous chondrites potentially experienced much more evaporation, although the evidence from Zn, Cd and Cu isotopes argues against evaporation. Unfortunately, the low Na contents also make the measurements of olivine phenocrysts much more challenging. The low concentrations make electron probe analyses impractical, and while in principle SIMS measurements do have the necessary sensitivity ubiquitous Na surface contamination of thin sections has so far hindered any systematic studies.

### 6.3.4 Iron Oxide and Metal

Iron is the next most volatile major element after Na and K, whether it is present as FeO or Fe-metal (Tachibana et al., 2015). SIMS measurements of Fe isotopic compositions of olivine phenocrysts in ordinary chondrite chondrules with a wide range of FeO contents reveal no systematic fractionations that would be consistent with Rayleigh behavior although they should have been detectible if present (Alexander and Wang, 2001). Since then, higher precision bulk chondrule measurements of chondrules from a number of chondrites have found only relatively small mass fractionation effects (Table 6.3). These mass fractionation effects are much smaller than would be predicted if their range of Fe contents were the result of free evaporation. Since chondrules are a major component of chondrites, if they experienced significant Fe evaporation one might expect to see the isotopic signatures even in bulk chondrite measurements. However, as with the chondrules, bulk chondrites exhibit very subdued Fe isotopic fractionations (Zhu et al., 2001).

Metallic iron is a common component of chondrules, particularly FeO-poor ones (type I). Texturally, Fe-metal seems to have been stable during chondrule formation. Villeneuve et al. (2015) showed by experiment that at high temperatures FeO-poor olivine (type IA-like) chondrules are readily oxidized to FeO-rich olivine (type IIA-like) chondrules, an argument for open-system addition requiring O-enriched vapor. However, Humayun et al. (2010, their Fig. 2) observed depletions of refractory siderophiles (e.g., W, Re, Os, Ir) relative to Fe, Co and Ni in metal grains in the rims of chondrules in CR chondrites, compared to their abundances in metal grains in the chondrule interiors. They argued for the recondensation of the metal in rims from a refractory-depleted vapor formed during chondrule melting. Close variants of this hypothesis have also been proposed by others for CR chondrites (Zanda et al., 1994; Kong et al., 1999; Connolly et al., 2001; Humayun et al., 2002; Campbell et al., 2005).

In contrast, Ebel et al. (2008, their Fig. 7) observed that interior metal grains in igneously zoned CR chondrules fall on the high temperature portions of Co/Fe vs. Ni/Fe trends predicted by condensation calculations that include a nonideal metal solution model (Ebel and Grossman, 2000), while exterior metal grains had solar values. In these igneously zoned CR chondrules, interior portions associated with high Co, Ni metal have nearly pure forsterite mineralogy, while outer parts are primarily enstatitic pyroxene (Hobart et al., 2015). They concluded that both silicate and metal chemistry are consistent with the melting of successive igneous rims of increasingly less refractory bulk compositions. In CR chondrites, stable melts with low FeO content in silicates require only modest enrichments in dust (Ebel and Grossman, 2000; Ebel, 2006, Plate 10). Similar major and trace element studies have yet to be conducted on meteorites from other chondrite groups.

### 6.3.5 Magnesium and Silicon

It is well documented that significant evaporation of Mg, Si and O occurred in many Type B (melted) CAIs in CV3 chondrites, primarily Allende (Grossman et al., 2000, 2002; Richter et al., 2002, 2007). The melting temperatures of Type B CAIs (1700-1800 K, e.g., Stolper, 1982; Stolper and Paque, 1986) are lower than those of FeO-poor Type I chondrules (1900-2100 K, e.g., Hewins and Radomsky, 1990; Alexander et al., 2008). Hence, these observations imply that if there was free evaporation of Mg, Si and O during chondrule formation (from chondrule melts) there should be clear evidence for it in the Mg, Si and O isotopes of chondrules.

Galy et al. (2000), Young et al. (2002) and Young and Galy (2004) showed that whole-chondrules from Allende (CV3.6), Bjürbole (L/LL4), and Chainpur (LL3.4) contain slight $^{25}$Mg excesses that in Allende correlate with chondrule Mg/Al and size. These relations are consistent with small amounts of evaporation during chondrule formation. Recently, Deng et al. (2017) have improved the precision for measuring Mg-Al systematics by laser ablation inductively coupled plasma mass spectrometry (LA ICP-MS) and confirmed for Allende and Leoville (CV3) chondrules the presence of only small levels of Mg isotopic mass fractionations. Similarly small mass fractionation effects have been reported for Si isotopes in chondrules (Molini-Velsko et al., 1986; Clayton et al., 1991; Georg et al., 2007; Hezel et al., 2010). As with Fe, there are only very small Si isotope effects evident in bulk chondrites (Georg et al., 2007; Fitoussi et al., 2009).

### 6.4 Implications of the Lack of Evidence for Evaporation

If the range of elemental abundances observed in chondrules and reflected in whole rock compositions were produced by free evaporation, they should be accompanied by large and systematic isotopic fractions that are characteristic of Rayleigh-type behavior. The fact that such fractionations are not seen in any of the isotope systems that have been studied has potentially profound implications for conditions during chondrule formation. In addition, despite a wide range of Na contents in Semarkona chondrules, at least, the Na appears to have remained at roughly the observed abundances throughout chondrule cooling irrespective of chondrule type. Indeed, Zn, Cd and Cu exhibit *light* isotope enrichments in chondrules relative to bulk meteorite. How can these observations be explained?

Galy et al. (2000) suggested that chondrules formed with high partial pressures (~1 bar) of $H_2$ present. High pressures of $H_2$ can reduce isotopic fractionation associated with evaporation both by enhancing evaporation rates relative to diffusion rates in the melt, and by restricting diffusion rates in the gas so that there is more back reaction between the melt and the evaporated material (Ebel 2006, plate 7). However, such high gas pressures are astrophysically unreasonable, making this explanation unlikely.

Another suggestion is that chondrules were heated and cooled very rapidly, thereby preventing significant evaporation (Wasson and Rubin, 2002). This explanation requires that the kinetics of melting and crystallization/solidification be much faster than those of evaporation. Rapid heating and cooling experiments are difficult to conduct, but Knudsen cell experiments (Imae and Isobe, 2017), and diffusion experiments do not support this explanation (Richter et al., 2011). However, there are natural 'experiments' that can be used to test this idea. Micrometeorites and cosmic spherules are heated during atmospheric entry to a range of peak temperatures and cooled in a matter of seconds (e.g., Love and Brownlee, 1991). Despite these very short heating times, they show very large elemental and isotopic fractionations in K, Fe, O, Si and Mg consistent with evaporative loss (Alexander et al., 2002; Taylor et al., 2005). The likely explanation is that the shorter the heating time the higher the degree of superheating required to achieve melting and, therefore, the faster the rates of evaporation. Thus, brief heating events are not the explanation for the lack of isotopic mass fractionation in chondrules.

### 6.4.2 High Solid Densities During Chondrule Formation

In the absence of plausible alternatives, the most likely explanation for the absence of systematic isotopic fractionations in chondrules that are consistent with evaporation is that when chondrules formed they were stable melts that approached equilibrium with their surrounding gas. To generate an equilibrium vapor, there must have been some evaporation from chondrules and any other material present (e.g., dust, refractory inclusions, etc.). So there could have been some initial isotopic fractionation that was later all but erased by subsequent gas-melt re-equilibration. The extent of evaporation needed to generate the equilibrium vapor and the time-scales needed to reach equilibrium will have depended on the density of solids (chondrule precursors, dust, etc.) present in the chondrule forming regions and the partial pressure of $H_2$ - the $H_2$ increases the equilibrium vapor pressures of many species. The lower the solid densities in the chondrule forming regions the greater degree of evaporation that is required to generate the equilibrium vapor and the longer it takes to erase any isotopic fractionations. Thus, the elemental and isotopic compositions of chondrules can potentially be used to constrain their formation conditions if the equilibrium and kinetic factors that would have controlled the evolution of their compositions are understood.

Equilibrium models indicate that to produce silicate melts, and to produce chondrule-like compositions by the time their solidi are reached, the solids in the formation regions must have been enriched in CI-like solids (i.e., dust) by factors of 15-100 (by numbers of atoms), relative to solar, at a total pressure of $P^{tot}=10^{-3}$ bars (Ebel and Grossman, 2000; Ebel, 2006, Plate 10). A total pressure of $10^{-3}$ bars is about the upper limit for astrophysical estimates of pressures in protoplanetary disks, even in shocks (D'Alessio et al., 2005). Higher enrichments are required for lower $P^{tot}$. Enrichments >1000 times solar even at $10^{-3}$ bars are needed if at near liquidus temperatures FeO-rich (type II) chondrules retained their current bulk FeO contents and metal melts were stable in type I chondrules (Ebel and Grossman, 2000; Alexander, 2004; Alexander and Ebel, 2012; Tenner et al., 2015). Kinetic models of chondrule evaporation and gas-melt re-equilibration also suggest that solids densities must have been >1000 at $10^{-3}$ bars or chondrule phenocrysts would have preserved isotopic mass fractionations in Fe, Si, O and Mg that are not observed (Alexander, 2004; Fedkin and Grossman, 2013).

However, the Na contents of chondrules provide the most stringent constraints. Even at 1000 times dust enrichments and $P^{tot}=10^{-3}$ bars, at near liquidus temperatures Na and K will be entirely in the gas, and only recondense into the melt after most olivine and pyroxene has already crystalized. This is because the vapor pressures of Na alone must be $> 10^{-3}$ bars for Na to condense into chondrules at near their liquidus temperatures (Lewis et al., 1993; Alexander and Ebel, 2012; Fedkin and Grossman, 2013). Thus, orders of magnitude higher solids enrichments are needed if chondrule melts maintained roughly their current Na contents at near liquidus temperatures (Alexander et al., 2008; Fedkin and Grossman, 2013). Furthermore, such solids enrichments would drive significant FeO condensation into chondrule silicates (Ebel and Grossman, 2000, their Figure 8), yet high Na abundances are observed even in FeO-poor (type I) chondrules (Alexander et al., 2008; Kropf and Pack, 2008; Kropf et al., 2009). While Na contents provide the most stringent constraints at the moment, maintaining the observed FeO and Fe-metal contents at near-liquidus temperatures also require very high solids densities, which is a particularly important constraint for carbonaceous chondrite chondrules for which it is unclear whether the alkalis were retained at near peak temperatures. Alexander and Ebel (2012) summarized the constraints that can be placed on chondrule densities during formation from Na and Fe abundances. They also placed lower limits on chondrule number densities during chondrule formation, requiring compound chondrule and nonspherical chondrule abundances that are significantly higher than previously estimated (Ciesla and Hood, 2004; Rubin and Wasson, 2005). Recent petrologic evidence for so-called cluster chondrites (Metzler, 2012; Metzler and Pack, 2016;

Bischoff et al., 2017) suggests that chondrule densities during formation may have been much higher than these lower limits, and more in line with the density estimates based on Na.

### 6.5  Outstanding Problems

A model of chondrules as stable melts that formed in regions that had been enormously enriched in solids, relative to solar, explains many features of chondrules. However, the solids enrichments or absolute densities that are required by Na retention during ordinary chondrite chondrule formation and Fe metal retention during ordinary and carbonaceous chondrite chondrule formation are much higher than current nebula models can explain (D'Alessio et al., 2005; Cuzzi et al., 2008). Also, to melt the chondrules in such dense clumps requires a formation mechanism that can focus a large amount of energy into small regions, and no nebular model has been shown to be able to do this. Given the potential astrophysical implications, it is important to continue to test such models.

There are, in fact, chemical and textural features that are not apparently entirely consistent with a high solids density model. For instance, as reviewed above, chondrules do retain small isotopic fractionations in many elements that are inconsistent with melts that fully equilibrated with their surrounding vapor at all stages of formation. These fractionations could simply reflect incomplete re-equilibration after the initial phase of evaporation, and/or incomplete equilibration between chondrules that had wide ranges of precursor compositions (Alexander and Ebel, 2012). The diversity of chondrule compositions indicate a wide range of precursors, and their diverse textures, even within single meteorites, require a broad range of thermal histories (Jones et al., 2005; Ebel et al., 2016).

Vapor from chondrules at the peripheries of a chondrule forming region would be able to diffuse out of the region before equilibrating with chondrules. As a result, chondrules at the peripheries of a region would be expected to show significant isotopic fractionations, yet few chondrules, if any, do. This suggests the possibility of crudely estimating the typical size of a chondrule forming region. Assuming that fewer than 1% of chondrules exhibit significant isotopic fractionation associated with evaporation, and a spherical formation region with a homogeneous density of chondrules, then estimates of diffusion distances in the gas for typical estimates of chondrule formation timescales and conditions require that such "particle-vapor clump" regions must have been at least 150-6000 km in radius (Cuzzi and Alexander, 2006). If they were any smaller, more than 1 % of the chondrules would have been within one diffusion distance of the surface of the region.

Chondrules even within a single chondrite exhibit a considerable diversity of compositions and textures that indicate a broad range of potential formation conditions (Jones et al., 2005). Perhaps most puzzling are compound chondrules composed of two or more distinct chondrule types that collided and stuck together while they were still hot and plastic. Since equilibration between chondrules involves diffusive exchange via the gas and diffusion distances in the gas on chondrule-forming timescales would have been much less than the sizes of the formation regions estimated above, it is possible that individual chondrule formation regions maintained numerous microenvironments that produced chondrules with different compositions (Cuzzi and Alexander, 2006). The development of turbulent mixing in such zoned chondrule forming regions is one possible explanation for multi-type compound chondrules. However, it also seems likely that the diversity of chondrule compositions and textures within chondrites requires multiple formation events with differing conditions.

### 6.5.1  Silica Metasomatism and the Sodium Paradox

Perhaps the most difficult observation to reconcile with formation of chondrules in dense clumps in which there was relatively little evaporation even of Na at near-liquidus temperatures is the presence of chondrules with pyroxene-rich outer zones around olivine-rich cores. The favored explanation for these rims is that volatilized SiO in the gas recondensed onto the chondrules creating silica enriched melts at their peripheries from which pyroxene crystallized. This process would seem to be entirely at odds with observations of Na and Fe metal being retained at near liquidus temperatures (Sections 6.3.3, 6.3.4), since Si is less volatile than Na or Fe.

Examples of mineralogically zoned chondrules with Na-bearing olivine in their centers and low-Ca pyroxene in their outer zones (with poikilitic textures, olivine enclosed in low-Ca pyroxene) have been highlighted by several petrological surveys of different chondrite types (Scott and Taylor, 1983; Tissandier et al., 2002; Krot et al., 2004; Jones, 2012; Libourel et al., 2006; Friend et al., 2016). The fraction of zoned chondrules in carbonaceous chondrites is estimated to be as high as 75% of all chondrules (Friend et al., 2016) demonstrating: 1) the importance of this petrologic feature, and 2) the need to take account of this fact in chondrule formation models.

Based on experiments that control the partial pressure of SiO gas above a crystallizing silicate melt (Tissandier et al., 2002; Kropf and Libourel, 2011), it has been suggested that gas-melt interactions may have played a major role during the formation of type I chondrules. According to these results, the formation of low-Ca pyroxene in a chondrule melt in which olivine is being dissolved:

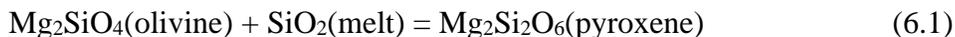

$$Mg_2SiO_4(olivine) + SiO_2(melt) = Mg_2Si_2O_6(pyroxene) \qquad (6.1)$$

is buffered by exchange with the surrounding nebular gas:

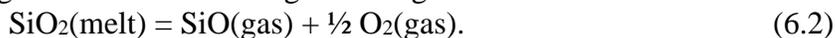

$$SiO_2(melt) = SiO(gas) + \tfrac{1}{2} O_2(gas). \qquad (6.2)$$

Increasing the activity of $SiO_2$ in the melt promotes crystallization of low-Ca pyroxene (or silica) if saturation conditions are reached in the chondrule melt (Reaction 6.1). In this scenario (Libourel et al., 2006), the chondrule melt is considered to be an open system and its composition, especially the $SiO_2$ content, is controlled by (1) exchange with the surrounding gas, controlled by the SiO gas partial pressure, $P_{SiO}(gas)$ (Reaction 6.2), (2) the dissolution of precursor olivines (Kropf and Libourel, 2011; Soulié et al., 2017), and (3) the crystallization of low-Ca pyroxenes (Tissandier et al., 2002; Chaussidon et al., 2008; Harju et al., 2015; Friend et al., 2016). The chemistry of the type I chondrules (PO, POP, PP) is thus dependent on $P_{SiO}(gas)$ and/or the duration of gas-melt interactions at high temperatures.

### 6.5.2  Sodium Zoning in Mesostasis

A separate problem is the presence of Na in chondrule mesostasis, increasing toward chondrule rims. Grossman et al. (2000, 2002), and Alexander and Grossman (2005) concluded that this Na zoning is due to secondary exchange with matrix in chondrite parent bodies, based on correlations between Na, Cl and H in mesostasis. They noted that some chondrules are not zoned because they have fully equilibrated with Na in matrix

Others have observed that mesostasis compositions enriched in silica have higher $Na_2O$ contents (Matsunami et al., 1993; Grossman et al., 2002; Libourel et al., 2003). Similarly, porphyritic pyroxene (PP) chondrules are systematically enriched in alkali and silica compared to porphyritic olivine pyroxene (POP) chondrules; porphyritic olivine (PO) chondrules being in comparison depleted in both Na and Si (Libourel et al. 2003; Berlin 2010). Sodium diffusion in molten silicate is four orders of magnitude faster than diffusion of Si. This work leaves open the possibility of Na influx into chondrules prior to parent body accretion.

Mathieu et al. (2008, 2011) have shown that Na solubility is primarily controlled by the silica content of the melt according to a bond species reaction of the type:

$$[1/2 \ Si - O - Si] + [1/2 \ Na - O - Na] = [Na - O - Si]. \qquad (6.3)$$

Addition of Na inhibits the polymerization of $[SiO_4]^{4+}$ tetrahedra in the melt. Reaction 6.3 clearly indicates that the more polymerized a melt the higher its Na solubility. Condensation of SiO into chondrule melts, i.e., increasing activity of $SiO_2$(melt) and hence increasing Si-O-Si linkages of the melt, would enhance Na solubility. The very low Si diffusivity would induce a zonation of $SiO_2$ in the melt, from rim to core. If the distribution of Na is controlled by the diffusion of $SiO_2$ in the chondrule melt according to reaction 6.3, then the heterogeneities in $Na_2O$ contents measured in chondrules could reflect variations of $P_{SiO}$(gas) and/or different timing of the gas-melt interaction. Higher $P_{SiO}$(gas) and/or longer exposure time would be required to produce PP chondrules. This scenario likewise could explain silica-bearing rims in CR chondrules (Krot et al., 2004), silica-bearing chondrules in enstatite chondrites (Lehner et al., 2013; Piani et al., 2016), and possibly silica-rich components in ordinary chondrites (Hezel et al., 2006).

In the context of the particle-vapor clump model (Cuzzi and Alexander, 2006; Section 6.5.1), this proposed phenomenological model suggests variations of vapor composition or cooling time. Although it is consistent with multiple separate chondrules reservoirs, which are subjected to various conditions over extended time periods (Jones, 2012), this scenario must be tested by a systematic survey of silicon isotope systematics in chondrules.

## 6.6 Conclusions

The recent decade has seen a revolution in non-traditional stable isotope investigations enabled by LA-ICPMS methods (e.g., Young and Galy, 2004; Moynier et al., 2017). These studies have produced little evidence of evaporation of chondrule melts, and puzzling evidence of chalcophile element light isotopic enrichments in chondrules. Measurements of Na in olivine phenocrysts have prompted new questions about chondrule melt - vapor interactions, suggesting high partial pressures of Na due to heating of regions that were highly enriched in solids. New experiments demonstrate how late SiO condensation into chondrules could explain their olivine/pyroxene petrology and perhaps also Na zoning in chondrule mesostases, but does not explain why Fe-metal or FeO do not similarly enter chondrules. In detail, distributions of Na, $SiO_2$ and FeO inside chondrules do not appear to fit prevailing models for chondrule formation. It might appear that with lots of data, there has been little progress (Wood, 2001). However, the abundance of new *kinds* of data bearing on vapor-melt exchange, while prompting new questions, also promises new, quantitative constraints on dynamical models for chondrule formation.

## Acknowledgements

The authors thank Sara Russell and Alexander Krot for comprehensive reviews. Research was funded by NASA Emerging Worlds grant NNXI6AD37G (DE). This research has made use of NASA's Astrophysics Data System.

Ebel et al., Vapor-Melt Exchange                            21